\documentstyle[12pt]{article}
\newcommand{\be}{\begin{equation}}
\newcommand{\ee}{\end{equation}}
\newcommand{\bea}{\begin{eqnarray}}
\newcommand{\eea}{\end{eqnarray}}
\newcommand{\bean}{\begin{eqnarray*}}
\newcommand{\eean}{\end{eqnarray*}}
\newcommand{\ba}{\begin{array}}
\newcommand{\ea}{\end{array}}

\newcommand{\slashl}[1]{\not{\!\!#1}}

\newcommand{\norsl}{\normalsize\sl}
\newcommand{\norsc}{\normalsize\sc}
\begin{document}

\begin{titlepage}

\title{QCD Corrections to the Nucleon's
       Spin Structure Function $g_2(x,Q^2)$}

\author{
\norsc  Jiro KODAIRA and Yoshiaki YASUI\thanks{Supported in part by
          the Monbusho Grant-in-Aid for Scientific Research
          No. 050076.} \\
\norsl  Dept. of Physics, Hiroshima University\\
\norsl  Higashi-Hiroshima 739, JAPAN\\
\\
\norsc  Kazuhiro TANAKA\\
\norsl  Dept. of Physics, Juntendo University\\
\norsl  Inba-gun, Chiba 270-16, JAPAN\\
\\
\norsc  Tsuneo UEMATSU\thanks{Supported in part by
          the Monbusho Grant-in-Aid for Scientific Research
          No. C-06640392.} \\
\norsl  Dept. of Fundamental Sciences\\
\norsl  FIHS, Kyoto University\\
\norsl  Kyoto 606-01, JAPAN\\}

\date{}
\maketitle

\begin{abstract}
{\normalsize
\noindent
We investigate the renormalization of the twist-3 operators
which are relevant for the  
spin-dependent structure function $g_{2}(x, Q^{2})$. We derive the
anomalous dimension for the non-singlet part by
calculating the off-shell Green's functions of the twist-3
operators including the operators which are proportional to the
equation of motion.} 
\end{abstract}

\begin{picture}(5,2)(-340,-620)
\put(2.3,-110){HUPD-9603}
\put(2.3,-125){JUPD-9611}
\put(2.3,-140){KUCP-94}
\put(2.3,-155){March, 1996}
\end{picture}
 
\vspace{2cm}
\leftline{\hspace{1cm}hep-ph/9603377}
 
\thispagestyle{empty}
\end{titlepage}
\setcounter{page}{1}
\baselineskip 24pt

The nucleon spin structure observed in the deep inelastic scattering
is described by the two 
structure functions $g_1(x,Q^2)$ and $g_2(x,Q^2)$. 
In the framework of the operator product expansion and the
renormalization group, not only the twist-2 operators
but also the twist-3 operators contribute to $g_2$ in the
leading order of $1/Q^2$ \cite{HM}. The general feature
characteristic to the higher-twist operators is the occurrence of
the operators which are proportional to the equation of motion
(EOM operators) \cite{POLI}. Analyzing the twist-3
operators must be interesting and useful from the theoretical viewpoint,
since they are the simplest
examples of the higher-twist operators.

There have been a lot of works \cite{JC,BKL,SVETAL} on the
$Q^{2}$-evolution of $g_{2}$.
We here mention the two works which are frequently cited.
(i) Ji and Chou \cite{JC} computed anomalous dimensions
of the twist-3 operators for $g_{2}$ in the Feynman gauge (for 
general spin $n$). They employed the 
massless on-shell scheme to 
compute the three-point function. However, it is not clear 
how the EOM operators are dealt with since their scheme itself might
not be consistent due to the infrared singularities
coming from the collinear configurations.
(ii) Bukhvostov, Kuraev and Lipatov\cite{BKL} 
derived GLAP-type evolution equation. But, this was carried out 
in the axial gauge and the relation to the covariant approach
is unclear.

The results of these two works seem not to be identical.
Thus, the computation of the anomalous dimensions 
in a covariant gauge in a fully consistent scheme is desirable.
To do this, we decided to compute the 
off-shell Green's functions to 
renormalize the operators. In this case, the EOM
operators should be included as 
independent operators. Infrared cut-off is provided by the external
off-shell momenta.
Recently this scheme has been employed in Ref.\cite{KYU} and
the complete calculation of the anomalous 
dimensions for the lowest ($n=3$) moment was demonstrated. 
The consistency and the efficiency of the method were also argued.
Here we extend the computation to the case for the general moment $n$.
We consider the flavor non-singlet case.

Phenomenologically, the first data of $g_{2}$ 
via the polarized deep inelastic scattering have been reported
recently \cite{ad}.
The extensive study will be performed in HERMES. 
The theoretical determination of the $Q^{2}$-dependence of $g_{2}$
is indispensable to extract the physical 
information from experimental data. 
Also, the $Q^{2}$-dependence itself will be checked and provide a deeper
test of QCD. In view of these, it is extremely important 
to establish the theoretical prediction, which is yet controversial as
discussed above.

We follow the convention of ref.\cite{KYU}. 
Let us first list up the gauge invariant twist-3 operators
which contribute to the moment $\int_{0}^{1}dx x^{n-1} g_{2}(x, Q^{2})$.
In the following expressions,
we suppress the flavor matrices $\lambda_i$ for the quark field
$\psi$.
\bea
  R_{n,F}^{\sigma\mu_{1}\cdots \mu_{n-1}} &=&
         \frac{i^{n-1}}{n} \Bigl[ (n-1) \overline{\psi}\gamma_5
       \gamma^{\sigma}D^{\{\mu_1} \cdots D^{\mu_{n-1}\}}\psi
           \nonumber\\
   & & \qquad - \sum_{l=1}^{n-1} \overline{\psi} \gamma_5
       \gamma^{\mu_l }D^{\{\sigma} D^{\mu_1} \cdots D^{\mu_{l-1}}
            D^{\mu_{l+1}} \cdots D^{\mu_{n-1}\}}
             \psi \Bigr] - {\rm traces} \ , \label{quark}\\
  R_{n,l}^{\sigma \mu_{1} \cdots \mu_{n-1}} &=& 
     \frac{1}{2n}\left(V_{l}-V_{n-1-l} + U_{l} + U_{n-1-l} \right)
     \;\;\;\;\;\;\;(l = 1, \cdots, n-2) \ , \label{eq:rl}\\
  R_{n,m}^{\sigma \mu_{1} \cdots \mu_{n-1}}&=&i^{n-2}{\cal S}m_{\psi}
          \overline{\psi}\gamma_{5}\gamma^{\sigma}
           D^{\mu_{1}}\cdots D^{\mu_{n-2}}\gamma^{\mu_{n-1}}\psi
             - {\rm traces} \ , \label{eq:rm}\\
  R_{n,E}^{\sigma \mu_{1} \cdots \mu_{n-1}} &=& i^{n-2}\frac{n-1}{2n}
   {\cal S} \left[ \bar{\psi}(i \slashl{D} - m_{\psi} ) 
  \gamma_5 \gamma^{\sigma} D^{\mu_{1}} \cdots D^{\mu_{n-2}}
        \gamma^{\mu_{n-1}} \psi \right.\nonumber \\
     & & \qquad + \,\left.\bar{\psi} \gamma_5 \gamma^{\sigma}
    D^{\mu_{1}} \cdots D^{\mu_{n-2}}\gamma^{\mu_{n-1}}(i\slashl{D}
      - m_{\psi} ) \psi \right] - {\rm traces} \ , \label{eq211}
\eea
where
\bean
   V_{l}&=&i^{n} g {\cal S} \bar{\psi}\gamma_{5} 
       D^{\mu_1} \cdots G^{\sigma \mu_l} \cdots
       D^{\mu_{n-2}}\gamma^{\mu_{n-1}} \psi - {\rm traces},\\
   U_{l}&=&i^{n-3} g {\cal S}\bar{\psi}
      D^{\mu_1} \cdots \tilde{G}^{\sigma \mu_l} \cdots
      D^{\mu_{n-2}}\gamma^{\mu_{n-1}} \psi - {\rm traces}.
\eean
In the above equations, (- traces) stands for the subtraction of
the trace terms to make the operators traceless and $D^{\mu}$ is
the covariant derivative.
$\{ \quad \}$ means symmetrization over the Lorentz indices,
${\cal S}$ the symmetrization over $\mu_i$ and $g$ the QCD
coupling constant.
$m_{\psi}$ represents the quark mass (matrix). 
The operators in (\ref{eq:rl}) contain
the gluon field strength $G_{\mu\nu}$ and the dual tensor
$\widetilde{G}_{\mu \nu}={1\over
2}\varepsilon_{\mu\nu\alpha\beta}
G^{\alpha\beta}$ explicitly.
The above operators are not all independent but satisfy the following
equation,
\be
 R_{n,F}^{\sigma\mu_{1}\cdots \mu_{n-1}} = 
        \frac{n-1}{n} R_{n,m}^{\sigma\mu_{1}\cdots \mu_{n-1}}
             + \sum_{l=1}^{n-2} (n-1-l)
                 R_{n,l}^{\sigma\mu_{1}\cdots \mu_{n-1}} +
             R_{n,E}^{\sigma\mu_{1}\cdots \mu_{n-1}} \ . \label{operel}
\ee
Therefore we can exclude one operator
among (\ref{quark})-(\ref{eq211}) to form an independent basis.
A convenient choice of the independent operators will be 
(\ref{eq:rl}), (\ref{eq:rm}) and (\ref{eq211}).
It is to be noticed that, for the $n$-th moment, $n$ gauge-invariant operators
participate in the renormalization.

We multiply the operators by the light-like vector
$\Delta_{\mu_{i}}$ to symmetrize the Lorentz indices and to
eliminate the trace terms. We then embed the operators
$\Delta \cdot R^{\sigma}_{n,l} \equiv \Delta_{\mu_{1}}\cdots
\Delta_{\mu_{n-1}} R_{n,l}^{\sigma\mu_{1}\cdots \mu_{n-1}}$
into the three-point function as
$\langle 0|{\rm T} \Delta \cdot R_{n,l}^{\sigma}(0) A_{\mu}(x)\psi(y)
\bar{\psi}(z)|0 \rangle$ and compute the
one-loop corrections.
We employ the Feynman gauge and
renormalize the operators in the MS scheme.
To perform the renormalization consistently,
we keep the quark and gluon external lines off-shell;
in this case, the EOM operator mix through the renormalization
as a nonzero operator.

One serious problem in the calculation is the mixing 
of the (many) gauge non-invariant EOM operators. As explained
in Ref.\cite{KYU}, these operators are given by replacing some of
the covariant derivatives $D^{\mu_i}$
by the ordinary derivatives $\partial^{\mu_i}$ in (\ref{eq211}).
This problem can be overcome by introducing the vector 
$\Omega_{\mu}$, which satisfies $\Delta^{\mu} 
\Omega_{\mu}=0$ \cite{kt}, and by contracting 
the Green's functions
$\langle 0|{\rm T} \Delta \cdot R_{n,l}^{\sigma}(0) A_{\mu}(x)\psi(y)
\bar{\psi}(z)|0 \rangle$
with this vector.
This brings the two merits: Firstly,
the tree vertices of the gauge invariant 
and non-invariant EOM operators coincide. Thus, essentially only one EOM
operator is now involved in the operator mixing. 
Secondly, the structure of the vertices for the twist-3 operators
are simplified extremely, and the computation becomes more tractable.

After the contraction with $\Omega_{\mu}$, the tree vertices
for $\Delta \cdot R^{\sigma}_{n,l}$ and 
$\Delta \cdot R^{\sigma}_{n,E}$ become respectively,
\be
  {\cal R}^{\sigma}_{n,l} = \frac{g}{2n} \hat{q}\left(
      (\hat{p}-\hat{q})^{n-2-l} \hat{p}^{l-1}
      \gamma^{\sigma}\slashl{\Omega} \gamma_5 \slashl{\Delta} -
    (\hat{p}-\hat{q})^{l-1} \hat{p}^{n-2-l}
   \slashl{\Omega} \gamma^{\sigma}\gamma_5 \slashl{\Delta} \right)t^a \ ,
                \label{eq304}
\ee
and
\be
  {\cal E}^{\sigma}_{n} = - \frac{n-1}{2n}g\left(
   (\hat{p}-\hat{q})^{n-2} \gamma^{\sigma} \slashl{\Omega}
     \gamma_5 \slashl{\Delta} + \hat{p}^{n-2}
   \slashl{\Omega} \gamma^{\sigma}\gamma_5 \slashl{\Delta} \right)t^a \ .
\ee
Here $p$ and $q$ are the momenta of the incoming quark and
the outgoing gluon
($\hat{p}\equiv\Delta\cdot p\,,\,\hat{q}\equiv\Delta\cdot q$), and $t^a$
is the color matrix normalized as
${\rm Tr}(t^at^b) = {1 \over 2}\delta^{ab}$. 

We present the results of the one-loop calculation
for the one-particle-irreducible three-point
function with the insertion of $\Delta \cdot R^{\sigma}_{n,l}$ (Fig. 1). 
We neglect the contribution of the operator
proportional to the quark mass, (\ref{eq:rm}), for the moment.
In the following expressions, 
$C_{F}= (N_{c}^{2}-1)/2N_{c}, C_{G}=N_{c}$, where
we assume $N_{c}$ color,
$S_{j} = \sum_{k=1}^{j} 1/k$, and 
$\epsilon = (4-D)/2$ with $D$ the space-time dimension.

Diagram (A) gives
\bea
 \lefteqn{\frac{g^{2}}{16 \pi^{2}\epsilon} (2C_{F} - C_{G})
  \left\{ \,\sum_{k=1}^{l-1} \frac{(-1)^{l-k} \ _{n-2}C_{k-1}}
   {(n-1)\ _{n-2}C_{l-1}} {\cal R}_{n,k}^{\sigma}
    + \frac{1}{n-1}{\cal R}_{n,l}^{\sigma}\right.}\nonumber\\
  & & \qquad\qquad\qquad\qquad\qquad\qquad\qquad\qquad\quad
   \left. + \sum_{k=l+1}^{n-2} \frac{(-1)^{l-k}\ _{n-2}C_{k}}
   {(n-1)\ _{n-2}C_{l}} {\cal R}_{n,k}^{\sigma} \right\}.
  \label{eq:eda}
\eea

Diagram (B) + diagram (C) gives
\bea
 \lefteqn{\frac{g^{2}}{16 \pi^{2} \epsilon}\,
 \left[ \, \sum_{k=1}^{l-1} \left\{ (2C_{F}-C_{G})
 \frac{(-1)^{l-k}\ _{l-1}C_{l-k}}{(n-l)\ _{n-k-1}C_{n-l}}
  - C_{G}\frac{n-2l-1}{(n-l)(l+1)} \right\} {\cal R}_{n,k}^{\sigma}
   \right.}\nonumber\\
 & & + \,\left\{ 2C_{F}(2 - S_{l}-S_{n-l-1}) -
  \frac{C_{G}}{l+1}\right\}{\cal R}_{n,l}^{\sigma} \nonumber \\
 & & \left. + (2C_{F}-C_{G}) \sum_{k=l+1}^{n-2}\frac{(-1)^{l-k}\
   _{n-l-2}C_{k-l}}{(l+1) \ _{k}C_{l+1}}{\cal R}_{n,k}^{\sigma}
   + C_{G} \frac{n-2l-1}{(n-1)(n-l)(l+1)}{\cal E}_{n}^{\sigma} \right].
    \label{eq:edb}
\eea

Diagram (D) + diagram (E) gives
\bea
   \lefteqn{\frac{g^{2}}{16\pi^{2}\epsilon} (2C_{F}-C_{G})
  \left\{ \sum_{k=1}^{l-1} \frac{2 (-1)^{k}\
   _lC_{k}}{l (l+1) (l+2)} {\cal R}_{n,k}^{\sigma}
     + \left( \frac{2 (-1)^{l}}{l (l+1)(l+2)} -
   \frac{(-1)^{l}}{n-l}\right) {\cal R}_{n,l}^{\sigma} \right.}
      \nonumber \\
 & & \qquad + \left. \sum_{k=l+1}^{n-2} \frac{(-1)^{n-k}\
   _{n-l-2}C_{n-k-2}}{n-l} {\cal R}_{n,k}^{\sigma}
   + \frac{1}{(n-1)(l+1)(l+2)}{\cal E}_{n}^{\sigma} \right\}.
   \label{eq:edc}
\eea

Diagram (F) + diagram (G) gives
\bea
 \lefteqn{\frac{g^{2}}{16 \pi^{2} \epsilon}C_{G}\left\{
   \sum_{k=1}^{l-1}\left( \frac{l+3}{2l(l+1)} +
    \frac{(l-2)(l-k+1)}{2l(l+1)(l+2)}-
  \frac{1}{n-l}\right) {\cal R}_{n,k}^{\sigma}\right.} \nonumber \\
 & & \qquad + \left(\frac{l+3}{2l(l+1)} + \frac{l-2}{2l(l+1)(l+2)}-
  \frac{1}{2(n-l)}\right){\cal R}_{n,l}^{\sigma} \nonumber \\
 & & \qquad \left.- \sum_{k=l+1}^{n-2}\frac{n-k-1}{2(n-l-1)(n-l)}
  {\cal R}_{n,k}^{\sigma} - \frac{n-2l-2}{(n-1)(n-l)(l+2)}
  {\cal E}_{n}^{\sigma} \right\}.
 \label{eq:edd}
\eea

Diagram (H) gives
\bea
   \lefteqn{\frac{g^{2}}{16\pi^{2}\epsilon}C_{G} \left\{
  \sum_{k=1}^{l-1}\left(\frac{1}{l-k}-\frac{1}{l}-\frac{k}{2l(l+1)}
  \right){\cal R}_{n,k}^{\sigma} \right.}\nonumber \\
 & & \qquad + \left(1 - S_{l} - S_{n-l-1} - \frac{1}{2(l+1)}
   - \frac{1}{2(n-l)}
   \right){\cal R}_{n,l}^{\sigma} \nonumber \\
 & & \qquad + \left. \sum_{k=l+1}^{n-2}\left( \frac{1}{k-l}-\frac{1}
   {n-l-1}
   - \frac{n-k-1}{2(n-l-1)(n-l)} \right){\cal R}_{n,k}^{\sigma} \right\}.
     \label{eq:ede}
\eea
The contributions from (\ref{eq:rm}) are easily calculated
by considering the quark two-point Green's function \cite{JC}.
The renormalization constants are determined in the MS scheme.
We summarize the final result in the following matrix
form:
\bea
\left(\matrix{R_{n,l}\cr
              R_{n,m}\cr
              R_{n,E}\cr}\right)_B =
\left(\matrix{Z_{lj}&Z_{lm}&Z_{lE}\cr
              0&Z_{mm}&0\cr
              0&0&Z_{EE}\cr}\right)
\left(\matrix{R_{n,j}\cr
              R_{n,m}\cr
              R_{n,E}\cr}\right)_R ,
\ \ \ \ \left(l,j = 1,\cdot\cdot\cdot,n-2\right),
\label{eq311}
\eea
where the suffix $R (B)$ denotes renormalized (bare) quantities.
We express $Z_{ij}$ as
\be
  Z_{ij} = \delta_{ij}+ {g^2 \over 16\pi^2 \varepsilon} X_{ij}
    \ \ \ \ \left(i,j= 1,\cdot\cdot\cdot,n-2, m , E\right).
  \label{eq312}
\ee
The relevant components of $X_{ij}$ are given as follows:
\bea
 X_{lj} &=&  C_G \frac{(j+1)(j+2)}{(l+1)(l+2)(l-j)} \nonumber \\
   & &+ \, (2C_F-C_G) \left[ (-1)^{l+j}\frac{\ _{n-2}C_{j-1}}
        {\ _{n-2}C_{l-1}  }
    \frac{(n-1+l-j)}{(n-1)(l-j)}+\frac{2(-1)^{j}}{l(l+1)(l+2)}
      \ _{l}C_{j}\right]\nonumber\\
    & & \ \ \ \ \ \ \ \ \ \ \ \ \ \ \ \ \ \ \ \ \ \ \ \ \ 
     (1\le j \le l-1), \label{eq313}\\
 X_{ll} &=& C_G \left(\frac{1}{l} - 
   \frac{1}{l+1}-\frac{1}{l+2}-\frac{1}{n-l}-S_{l}
    -S_{n-l-1} \right)\nonumber \\
   & & \qquad\qquad + (2C_F - C_G)\left[\frac{1}{n-1} 
   + \frac{2(-1)^{l}}{l (l+1)(l+2)}-\frac{(-1)^{l}}{n-l}\right]
           \nonumber \\
   & & \qquad\qquad\qquad\qquad\qquad\qquad
      + C_F \left(3-2S_{l} -2S_{n-l-1} \right),  \label{eq314}\\
 X_{lj} &=& C_G\frac{(n-1-j)(n-j)}{(n-1-l)(n-l)(j-l)} \nonumber\\ 
   & & + (2C_F - C_G) \left[ (-1)^{l+j}\frac{\ _{n-2}C_{j}}
     {\ _{n-2}C_{l}} \frac{(n-1-l+j)}{(n-1)(j-l)}
   +(-1)^{n-j}\frac{\ _{n-2-l}C_{n-2-j}}{n-l} \right] \nonumber\\
  & & \ \ \ \ \ \ \ \ \ \ \ \ \ \ \ \ \ \ \ \ \
    \left( l+1 \le j \le n-2\right), \label{eq315}\\
  X_{lm} &=& \frac{4C_{F}}{n l(l+1)(l+2)} \ ,\, 
 X_{lE} = \frac{2 C_{F}}{(n-1)(l+1)(l+2)} \,,\,
  X_{mm}= - 4C_{F}S_{n-1} \ . \label{eq316}
\eea
For the physical quantity (moments), only
the $Z_{lj}\,,\,Z_{lm}\,,\,Z_{mm}$ components of the
renormalization matrix give the contributions since the
physical matrix element of the EOM operators turns out to be zero.
With the above $X_{ij}$,
the anomalous dimension matrix for the twist-3 operators
becomes,
\be
  \gamma_{ij} = - {g^2 \over 8\pi^2 } X_{ij} \ .
         \label{eq321}
\ee
This matrix enters into the renormalization group equation for
the Wilson's coefficient function $E_i$ associated with the
corresponding operator as,
\[ \left[ \left( \mu \frac{\partial}{\partial \mu}
    + \beta (g) \frac{\partial}{\partial g} 
    - \gamma_m m \frac{\partial}{\partial m} \right) \delta_{ij}
    - \gamma_{ij} \right] E_i = 0 \ .\]

In the present study, we have obtained the anomalous
dimension for the twist-3 operators which contribute
to $g_2 (x,Q^2)$. We performed the calculation with the
manifest Lorentz covariance being kept. We adopted the
Feynman gauge and dimensional regularization.
We have chosen the operators which include the gluon field
strength explicitly as an independent operator's basis although
this choice of basis is never compulsory.
To identify the renormalization constants correctly,
the off-shell Green's functions are considered.
We have shown that the EOM operators
play an important role to complete
the renormalization of composite operators and the structure
of the renormalization constant matrix takes the triangular
form expected from the general argument \cite{COLL}. 

If we could calculate the \lq\lq on-shell\rq\rq matrix elements
of composite operators in terms of purely perturbative Feynman
graphs, we could obtain the enough informations without
considering the EOM operators. However the infrared singularity coming
from the collinear configuration can not be regulated \cite{EFP}.
Therefore we believe that calculating the off-shell Green's functions
is the safest method to obtain the anomalous dimensions.

As a technical issue, we have used the projection introduced in
Ref.\cite{kt} to avoid the complexity stemmed from the fact that
there are a lot of EOM operators which are not gauge invariant.
This projection may have some relation to the case in which
the light-cone gauge fixing $\Delta^{\mu} A_{\mu} = 0$ is
adopted since this gauge does not discriminate between
the covariant and the partial derivative in the composite
operators. 

Since the authors of Ref.\cite{JC} adopted the on-shell scheme to
extract the anomalous dimensions,
we can not compare our results with theirs graph by graph.
However our final results are not in agreement
with theirs although our definition of composite
operators and choice of gauge
are the same as those in Ref.\cite{JC}. If one replaces the index $i$ by
$n-1-i$ in their results, the both results become the same.

On the other hand, our final results agree with those of Ref.\cite{BKL}.
The approach adopted by them is quite different from ours.
The main differences are : (a) they considered the physical quantities
(structure functions) from the beginning, so
the EOM operators do not appear explicitly, (b) they adopted
the axial gauge to derive the GLAP-type evolution equation.
In their analyses, the evolution of the quark bilinear operator
(\ref{quark}) was also considered.
Therefore it seems that their choice of the operator basis
is different from ours at the intermediate stage of calculations.
By including the quark bilinear operator (\ref{quark}), we can
eliminate the gauge invariant EOM operator from the independent
operator basis (see (\ref{operel})).
Furthermore, they included some
contributions from the \lq\lq one-particle-reducible\rq\rq diagrams
to take into account the circumstance that the partons can not
be regarded as located on the mass shell \cite{BKL}.
This may correspond to the fact that even if the gauge invariant
EOM operator is discarded,
the gauge non-invariant EOM operators turn out to mix through the
renormalization \cite{KYU}.

We expect that future precise measurements on $g_2$ will
clarify the effect of twist-3 operators which may be
the first quantity to see the higher-twist effect in QCD.

\vspace{3cm}
We would like to thank H. Kawamura and Y. Koike for valuable discussions.

\vspace{2cm}
\centerline{\bf Figure Captions}

\vskip 15pt
\begin{description}
\item[Fig. 1] One-loop corrections
to $\Delta\cdot R_{n,l}^{\sigma}$. 
\end{description}

\newpage
\input epsf.sty
\pagestyle{empty}
\begin{figure}
\epsfxsize=17cm
\epsfysize=22cm
\epsffile{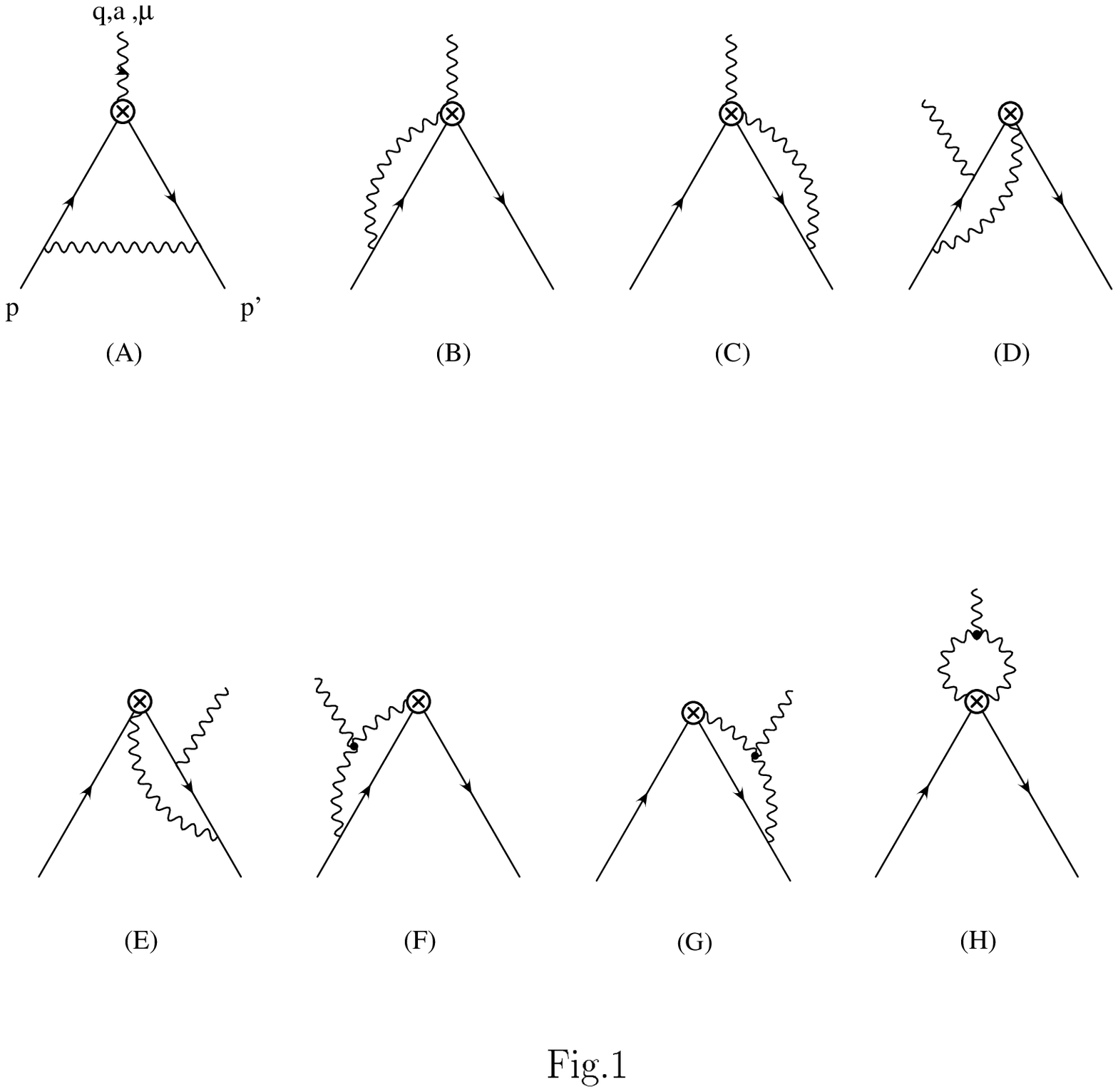}
\end{figure}

\end{document}